\documentclass[sigconf]{acmart}
\AtBeginDocument{%
  \providecommand\BibTeX{{%
    \normalfont B\kern-0.5em{\scshape i\kern-0.25em b}\kern-0.8em\TeX}}}

\copyrightyear{2021}
\acmYear{2021}
\setcopyright{acmcopyright}\acmConference[ArXiv]{Preprint}{2021}{ACM}

\usepackage{alltt	
          , multirow   
          , booktabs   
          , listings   
          , graphicx   
          , latexsym   
	, mathtools
	,verbatim
	,threeparttable
          , amsmath}    
\usepackage{balance}
\usepackage[all]{xy}          
\usepackage[perpage,bottom,stable,multiple]{footmisc}	
\usepackage{hhline,tabu}
\usepackage{makecell}
\usepackage{microtype}

\usepackage{algpseudocode}
\usepackage{algorithm2e}
\usepackage{tabularx,ragged2e,booktabs, caption, adjustbox}
\usepackage{subcaption}
\usepackage{makecell}
\usepackage{pgfplots}
\usepackage{bbold,bm}
 \usepackage{wrapfig}
 \usepackage{tcolorbox}
\usetikzlibrary{patterns}
\usepackage{xcolor,colortbl}
\usepgfplotslibrary{statistics}
\usetikzlibrary{positioning,fit,calc}
\pgfplotsset{
    box plot/.style={
        /pgfplots/.cd,
        black,
        only marks,
        mark=-,
        mark size=1em,
        /pgfplots/error bars/.cd,
        y dir=plus,
        y explicit,
    },
    box plot box/.style={
        /pgfplots/error bars/draw error bar/.code 2 args={%
            \draw  ##1 -- ++(1em,0pt) |- ##2 -- ++(-1em,0pt) |- ##1 -- cycle;
        },
        /pgfplots/table/.cd,
        y index=2,
        y error expr={\thisrowno{3}-\thisrowno{2}},
        /pgfplots/box plot
    },
    box plot top whisker/.style={
        /pgfplots/error bars/draw error bar/.code 2 args={%
            \pgfkeysgetvalue{/pgfplots/error bars/error mark}%
            {\pgfplotserrorbarsmark}%
            \pgfkeysgetvalue{/pgfplots/error bars/error mark options}%
            {\pgfplotserrorbarsmarkopts}%
            \path ##1 -- ##2;
        },
        /pgfplots/table/.cd,
        y index=4,
        y error expr={\thisrowno{2}-\thisrowno{4}},
        /pgfplots/box plot
    },
    box plot bottom whisker/.style={
        /pgfplots/error bars/draw error bar/.code 2 args={%
            \pgfkeysgetvalue{/pgfplots/error bars/error mark}%
            {\pgfplotserrorbarsmark}%
            \pgfkeysgetvalue{/pgfplots/error bars/error mark options}%
            {\pgfplotserrorbarsmarkopts}%
            \path ##1 -- ##2;
        },
        /pgfplots/table/.cd,
        y index=5,
        y error expr={\thisrowno{3}-\thisrowno{5}},
        /pgfplots/box plot
    },
    box plot median/.style={
        /pgfplots/box plot
    },
    boxplot/draw/median/.code={
   \draw [/pgfplots/boxplot/every median/.try]
       (boxplot box cs:\pgfplotsboxplotvalue{median},0)
         node[left] {\pgfmathprintnumber{\pgfplotsboxplotvalue{median}}}
        --
        (boxplot box cs:\pgfplotsboxplotvalue{median},1);
  },
}
\makeatletter
\usepackage[utf8]{inputenc}
\usetikzlibrary{arrows,automata}
\usetikzlibrary{shapes.geometric}
\tikzset{
    state/.style={
           rectangle,
           rounded corners,
           draw=black, thick,
           minimum height=2em,
           inner sep=0pt,
              text centered,
           },
    ball/.style = {circle, draw, align=center, anchor=north, inner sep=0},
    database/.style={cylinder,aspect=0.5,draw,rotate=90,path picture={
        \draw (path picture bounding box.160) to[out=180,in=180] (path picture bounding box.20);}},
    header node/.style = {
        Minimum Width = header nodes,
        font          = \ttfamily,
        text depth    = +0pt,
        fill          = white,
        draw},
    header/.style = {%
    inner ysep = +.7em,
    append after command = {
      \pgfextra{\let\TikZlastnode\tikzlastnode}
      node [header node] (header-\TikZlastnode) at (\TikZlastnode.north) {#1}
      node [span = (\TikZlastnode)(header-\TikZlastnode)]
        at (fit bounding box) (h-\TikZlastnode) {}
    }
  },
}

\newcommand*\circled[2][1.6]{\tikz[baseline=(char.base)]{
    \node[shape=circle, draw, inner sep=1pt, 
        minimum height={\f@size*#1},] (char) {\vphantom{WAH1g}#2};}}
\makeatother

\DeclareUnicodeCharacter{202F}{FIX ME!!!!}

\definecolor{pblue}{rgb}{0.13,0.13,1}
\definecolor{pgreen}{rgb}{0,0.5,0}
\definecolor{pred}{rgb}{0.9,0,0}
\definecolor{pgrey}{rgb}{0.46,0.45,0.48}
\definecolor{light-gray}{gray}{0.95}
\begin{document}

\title{Semantic Slicing of Architectural Change Commits: Towards Semantic Design Review}

\author{Amit Kumar Mondal}
\affiliation{%
  \institution{University of Saskatchewan}
\country{Canada}}
\email{amit.mondal@usask.ca}

\author{Chanchal K. Roy}
\affiliation{%
  \institution{University of Saskatchewan}
\country{Canada}}
\email{chanchal.roy@usask.ca}

\author{Kevin A. Schneider}
\affiliation{%
  \institution{University of Saskatchewan}
\country{Canada}}
\email{kevin.schneider@usask.ca}

\author{Banani Roy}
\affiliation{%
  \institution{University of Saskatchewan}
\country{Canada}}
\email{banani.roy@usask.ca}

\author{Sristy Sumana Nath }
\affiliation{%
  \institution{University of Saskatchewan}
\country{Canada}}
\email{sristy.sumana@usask.ca}

\renewcommand{\shortauthors}{Amit Kumar Mondal et al.}
\renewcommand{\shorttitle}{Semantic Slicing of Architectural Change Commits}


\begin{abstract}
Software architectural changes involve more than one module or component and are complex to analyze compared to local code changes. Development teams aiming to review architectural aspects (design) of a change commit consider many essential scenarios such as access rules and restrictions on usage of program entities across modules. Moreover, design review is essential when proper architectural formulations are paramount for developing and deploying a system. 
Untangling architectural changes, recovering semantic design, and producing design notes are the crucial tasks of the design review process. To support these tasks, we construct a lightweight tool \cite{archslice} that can detect and decompose semantic slices of a commit containing architectural instances. A semantic slice consists of a description of relational information of involved modules, their classes, methods and connected modules in a change instance, which is easy to understand to a reviewer. We extract various directory and naming structures (DANS) properties from the source code for developing our tool. Utilizing the DANS properties, our tool first detects architectural change instances based on our defined metric and then decomposes the slices (based on string processing). Our preliminary investigation with ten open-source projects (developed in Java and Kotlin) reveals that the DANS properties produce highly reliable precision and recall (93-100\%) for detecting and generating architectural slices. Our proposed tool will serve as the preliminary approach for the semantic design recovery and design summary generation for the project releases.
\end{abstract}

\ccsdesc[500]{Software and its engineering~Software creation and management}
\ccsdesc[300]{Software and its engineering~Software post-development issues}
\ccsdesc{Software and its engineering~Software evolution}

\keywords{Architectural change, semantic slice, design review}


\maketitle

\section{Introduction}\label{intro}
Software maintenance activities induce the most efforts and costs in a software system’s lifecycle \cite{le2015empirical_change}. Understanding and updating a system’s architecture is
crucial for software maintenance \cite{garcia2021constructing}. 
As parts of the maintenance activities, development teams review changes \cite{tang2014archi_review, caulo2020knowledge_review}, and design structure after completing a certain milestone or release (or in the late-lifecycle) for various reasons such as paying technical debts, fixing flaws or correcting the design structure \cite{alves2016technical_debt,ghorbani2019repair_javaarchi, kehrer2019jpms_container}. 
In this review process, requirements and associated design information extraction, semantic design recovery (concerned with the production of meaning, and how logic and language are used in designing a software \cite{hewitt2019semantic}), design summary generation, untangling changes, etc. are essential tasks \cite{tang2014archi_review, schmitt2020arcade, dias2015untangling}. Collectively, we refer to them as design review tasks \cite{tang2014archi_review}.
\par
For these tasks, architectural change instances are required to be detected and decomposed by the automated tools to reduce human efforts \cite{le2015empirical_change,ahmad2012change_knowledge}. In this paper, to support the design review tasks, we propose a lightweight tool for detecting module-level architectural changes and generating semantic slices of the change instances from source code. We rely on the directory and naming structures (DANS) properties for developing our tool. Please note that our extracted semantic slices are based on the structural relations that are mostly different than the slices generated by the existing tools \cite{dias2015untangling, li2017semantic, wang2019cora, li2019precise, wang2021can}. A structural semantic slice (SSC) consists of a description of relational information of involved modules, their classes, methods and connected modules in a change instance, which is easy to understand to a reviewer \cite{li2017semantic}. With these SSCs, various layers of abstract design summary generation are possible \citep{jiang2017change_summary}. For example, a high-level abstract summary can be automatically generated from such a slice as -- \emph{The ASBC \underline{module defines} a sensitive STC \underline{method with new} runtime \underline{dependency on ASC module} for authorizing sensitive service access} (underline texts represent an architectural relation, acronyms are discussed in Section \ref{jpms_op}). Besides, this information is crucial to feed into other DevOps tools such as design decision recovery \cite{shahbazian2018decision_recovery} and design note generation \cite{amitscam_2019}. 
\par
In a release or a milestone, the changes that contain architectural instances require special attention from the development team due to the far-reaching consequences \cite{paixao2019review_change, ghorbani2019repair_javaarchi}. However, software architectural changes involve in more than one module or component and are complex to analyze compared to local code changes. Architectural aspects (design) of a change task are reviewed considering crucial scenarios such as access rules and restrictions on usage of program entities across the modules. Moreover, design review is essential when proper architectural formulations are paramount for developing and deploying a system \cite{mak2017book_java9, kehrer2019jpms_container,ghorbani2019repair_javaarchi}. Researchers are working for supportive tools and techniques for reviewing change commits focusing on atomic or local changes \cite{dias2015untangling, wang2019cora}. 
However, little or no effort is given to support reviewing architectural change instances of those projects by slicing the structural relations. 
\par
As a preliminary study, in this paper, we investigate the SSC generation for Java Platform Module System (JPMS) based projects (in Java and Kotlin) \cite{nicolai2018java9_report, mak2017book_java9} (where the architectural organization is predominant). JPMS is introduced to handle coupling and dependency among modules to reduce bloated software, performance issue, security backdoors, and higher maintenance costs \cite{ghorbani2019repair_javaarchi, kehrer2019jpms_container} by well-defined modules and access restrictions among them. However, despite built-in supports for handling JPMS, development and maintenance teams require tracking a complex knowledge-base of static and run-time architecture and are still error-prone. That is why supportive review tools are urgently needed.

For our tool, we define an M2M (module to module) metric for high-level architectural change \cite{le2015empirical_change,garcia2013c2c} of JPMS based projects following the existing metrics. We identify several observations (Table \ref{extraction_info}) on the DANS properties of the program entities for detecting architectural change instances based on this metric and decomposing the SSCs. We extract the DANS properties (such as addition, deletion, moving or shrinking multiple class imports) based on regular expressions of string matching. Since our tool does not process the Abstract Syntax Trees (AST) properties, it does not require compiling each version (that requires human intervention to fix 3rd party library dependencies). Thus, our approach is easy to deploy with the version control system (VCS) APIs. Preliminary evaluation of our tool with ten open-source projects indicates that the DANS properties produce highly reliable precision and recall rate (93-100\%) for detecting and decomposing architectural change slices.
In summary, the key contributions of this paper are:

\begin{itemize}
    \item We construct a benchmark dataset containing module-level architectural change (M2M) commits and semantic slices of the changed code for advancing research in this domain. 
    \item We manually explore change commits and identify 16 types of directory and naming structure (DANS) properties necessary for automatic processing (for any tool) of the architectural change instances from the source code.
    \item We develop a lightweight tool using the DANS properties for module-level architectural change detection and semantic slice generation of the changes based on the special structural relations in the source code.
\end{itemize}
The rest of the paper is organized as follows. Section \ref{related_work} discusses related work. Section 3 presents the background and a motivating example. Section 4 presents the dataset collection process. Section 5 discusses the change detection process. Section 6 reports the semantic slice generation process. Section 7 presents performance results. Section 8 concludes our study with future work.
\vspace{-0.2cm}
\section{Related Work} \label{related_work}
\textbf{Architectural Change Detection and Design Decision Recovery:}
Software architecture can be defined into three levels of abstraction according to the convenience of the development team: (i) high-level -- where design models or modules are considered, (ii) intermediate level -- package, and classes are considered, and (iii) low level -- methods and functions are considered. 
MoJo \cite{tzerpos1999mojo}, MoJoFM \cite{wen2004mojofm}, $A\Delta$ \cite{jansen2008architectural_delta}, CB \cite{bouwers2011archi_analizability}, C2C \cite{garcia2013c2c}, and A2A \cite{le2015empirical_change} are focused on high-level change detection. In recent times, a few studies focus on recovering architectural design from the release history of software \cite{garcia2013c2c, shahbazian2018decision_recovery, schmitt2020arcade}. EVA \cite{nam2018eva} and ARCADE \cite{schmitt2020arcade} are excellent tools for recovering a static architecture and detecting changes based on ACDC/MoJo \cite{tzerpos2000accd} and ARC \cite{garcia2013c2c} techniques. However, these tools are explicitly dependent on other techniques to extract models and clusters (they are arbitrary and have no formal limit). Only expert intervention can ensure architectural change detection's accuracy of these metrics, and analysis of thousands of change versions is almost infeasible. Moreover, they cannot recover semantic design \cite{hewitt2019semantic}. We propose a module-level architectural change detection tool based on the developer's defined modules and thus more concrete and reliable. Our tool also extract semantic change relations on the detected change instances. Hence, our tool and benchmark data can be used to further validate and enhance the existing metrics for architectural change detection (available at \cite{archslice}).
\par
\textbf{Change Slicing for Code review:}
Here, we discuss a few of the most famous works among the existing studies \cite{dias2015untangling, li2017semantic, wang2019cora} for slicing the committed code. Dias et al. \cite{dias2015untangling} worked on tangled code change information slicing at the fine-grain statement level (i.e., one variable is associated with two lines of code, two files are changed together, the distance between two modified lines in a file, etc.) using AST properties to separate multiple intentions within a single commit. Later, they attempt to cluster the slices based on the pair relation, such as two methods are only refactored, two classes within the same package, etc. Li et al. \cite{li2017semantic} separate all types of atomic changes with the AST algorithm of a set of related commits in the version history for commit porting. Wang et al. \cite{wang2019cora} developed a more intelligent tool for decomposing changed code within a commit using AST parsing and machine learning. They cluster the code based on the class-level, method-level, field-level, and statement-level changes. Then they rank those changes considering the number of referenced variables for the code reviewers. Several studies have been enhancing atomic code change slicing works \cite{li2019precise, wang2021can}. However, these studies do not focus on architectural semantics and relations. Therefore, decomposing architectural change of a commit would enhance these techniques for reviewing more complex scenarios for architecture intensive systems. To reduce the gap, we attempt to generate architectural change slices for design review. 

\section{Motivation and Background} \label{jpms_op}
\textbf{Background:} 
This section aims to provide an idea of how architecture can be a centrepiece of modern-day software development and why reviewing architectural relations is crucial. From Java 9, we can define a concrete module compared to a conceptual module than we could before JPMS. Concrete modules accelerate containerization of cloud-service based systems more efficiently and securely \cite{kehrer2019jpms_container}. JPMS is built for supporting the following core principals \cite{mak2017book_java9}: (i) prevents unwanted coupling between modules, (ii) only exposes well-defined and stable interfaces to other modules, (iii) provides a reliable configuration of the dependent module, and (iv) controls reflective access to sensitive internal classes. The Java module system will have a profound impact on software development. In JPMS, a \emph{module} is a uniquely named collection of reusable packages which is defined by a descriptor file called \emph{module-info.java} having meta-data, including the declaration of named module \cite{ghorbani2019repair_javaarchi, nicolai2018java9_report}. A named module specifies (1) its dependencies of classes and interfaces (entities) on other modules and should specify (2) which of its entities are exposed to other modules for usage. Some of the operations provided by JPMS \cite{mak2017book_java9} for handling these specifications are: 
\begin{itemize}
\small
    \item \emph{requires (R)} - express its dependency on the other module. 
    \item \emph{provides (P)} - provides an implementation of an interface with another class as an implementation class. 
    \item \emph{opens/open (O)} - gives run-time access and open for use it with reflections. It is used to expose the whole module.
    \item \emph{uses (U)} - instructs run time loading of services. 
    \item \emph{transitive (T)} - expresses implied dependency for API. It ensures that any module which requires second module also implicitly requires third module (linked to the second module).
\end{itemize}
Each of these operations are architectural and have greater implication in terms of both static and run-time behaviors of a project. We call these operations as module operations ($MO$). 

\begin{figure*}
\includegraphics{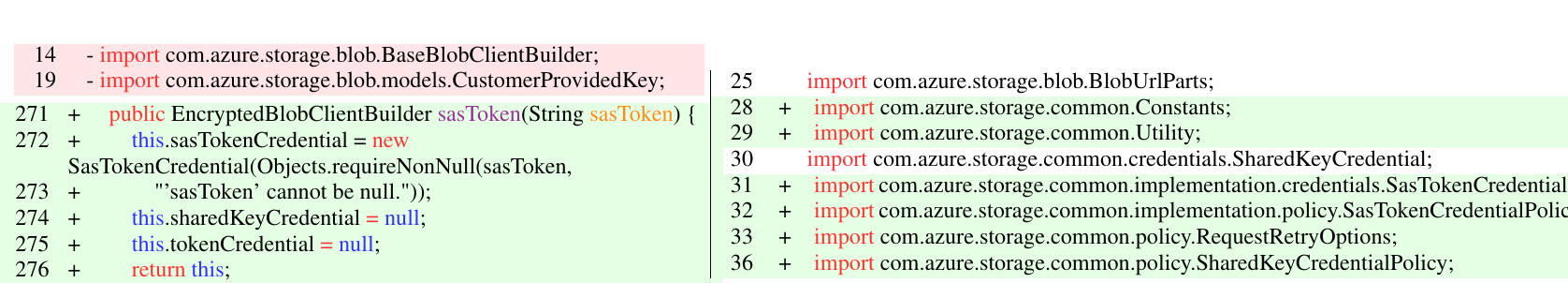}
 	\caption{A change that sets the SAS token used to authorize requests sent to an Azure service. Here, the plus (+) sign with green background indicates addition and the minus (-) sign with a red background indicates deletions. }
 		\label{example_change}
 		\vspace{-.4cm}
\end{figure*}

\textbf{Motivating Example:}
In this section, we explain a subset of modifications of \emph{EncryptedBlobClientBuilder (EBCB)} class of \emph{azure-storage-blob-cryptography (ASBC)} module from a commit of the Azure Java SDK project \cite{azure_commit}. The modifications are shown in Figure \ref{example_change} following the same GUI representation in github.com. 
In JPMS projects, usage of module entities has purposeful rules and restrictions due to runtime access of sensitive parts, dynamic containerization and separation of concerns. If the reviewers want to review architectural changes to revisit those, they must first identify whether the commit contains such changes. They also need to extract various other code related information \cite{li2017semantic, wang2019cora}. 
\par
For example, in Figure 1, a new method called \emph{sasToken()} is added in lines 271-276. The reviewers have to figure out which classes and modules are involved with this method and with which it has new dependencies. Variables in lines 272, 274 and 275 need to be searched in multiple places and compared for references to find the dependencies. Some of the candidates of those variations are discussed later in Section \ref{slicing}. However, statement 272 uses a newly imported class \emph{SasTokenCredential (STC)} in line 31. Next, they have to find which module it belongs to. But, the reviewers cannot determine the module with this GUI interface (or with the provided information by the VCS API) perfectly; not even if it is from the same module of the \emph{EBCB} class such as the \emph{BaseBlobClientBuilder} class in line 14. For instance, despite being from different modules, both lines 14 and 29 contain a significant portion of the common structure (\emph{com.azure.storage}). For these, they have to search the location of the class in the codebase (perhaps with the IDE). This manual search returns the directory from where they identify the module, which is the \emph{azure-storage-common (ASC)} module in this case (a cross-module). Such an instance is a candidate for revisiting the cross-module rules and restrictions due to various concerns discussed in the Background section. However, the commit contains modifications of 32 classes with many variations. Thus, the manual process to extract the crucial code information is time-consuming, tedious, and error-prone (such as the same class name exists within multiple modules). Besides, extracting the described information is crucial for many other analytic techniques such as design decision recovery and multiple intentions detection.
Consequently, our study contains two steps: (i) detecting commits containing architectural change instance, and (ii) then slicing those commits.

\section{Dataset Preparation}
We collect and prepare our experimental dataset in various phases. For collecting JPMS based projects, first, we search commits containing \textit{module-info.java} in GitHub. Thus, we get almost 200 projects, many of them are large or small or toy projects. Among them, we selected ten projects of various domains having the highest number of commits (and multiple modules), excluding native JDK-related projects. We filter out the commits having structural code change \cite{lutellier2015archi_recovercompare, amitscam_2019} or having modification in \emph{module-info.java} from all the commits in the period of July 2017 to July 2020 because of the official release of JPMS in 2017. In this way, we collect a total of 3,647 commits (\emph{Selected} column in Table \ref{table:dataset}). In the final phase, we manually determine the commits having an architectural change instance (M2M). It took around \textbf{240} working hours to complete the analysis of the commits. We found around 2,720 such commits, which are presented in the \emph{M2M} column in Table \ref{table:dataset}. We use this dataset for investigating the automated tool development utilizing the DANS properties.

\begin{table}[h]
\small
	\caption{M2M dataset and change detection performance.}\label{table:dataset}
	\vspace{-.2cm}
	\begin{center}
		\begin{tabular}{|p{.85in}|p{0.38in}|p{.38in} |p{.30
		in}| p{.17in} |p{.17in}|} \hline
			\textbf{Project} & \textbf{Commit} & \textbf{Selected} & \textbf{M2M} & \textbf{P}& \textbf{R}\\\hline \hline
			HibernateSearch\cite{hibernate_search}  & 9504 &53 &20&1.0& 1.0\\\hline
			Aion\cite{aion}  & 4718 & 1064&863&1.0& 0.98\\\hline
		    Webfx\cite{webfx}  & 3770 &  778& 563&1.0& 0.99\\\hline
			Speedment\cite{speedment}  & 4483 &  243& 222&0.97& 1.0\\\hline
			AzureSDK\cite{azure_sdk} & 15,180 & 276& 244&0.99& 1.0\\\hline 
			Atrium(Kotlin)\cite{atrium}& 1988 & 379&210& 0.98& 1.0 \\\hline
			Bach\cite{bach}  & 2114 & 365& 145&0.99&1.0\\\hline
		    Vooga\cite{vooga}  & 1210 &  447& 416&0.96& 1.0\\\hline
			Imgui(Kotlin)\cite{imgui} & 1703 &  35& 34&1.0& 1.0\\\hline
 			MvvmFX\cite{mvvmfx} & 1100 &  7& 3&NA&NA\\\hline 
             & Total  & 3647 & 2720 &&\\\hline
		\end{tabular}
	\end{center}
		\vspace{-.4cm}
\end{table}

\section{Architectural Change Detection} \label{change_detection}
\subsection{M2M Change Metric}
Detecting architectural change is an ongoing research. The most popular metrics for detecting higher level changes are $A\Delta$ \cite{jansen2008architectural_delta}, A2A \cite{le2015empirical_change}, MoJoFM \cite{wen2004mojofm}, and C2C \cite{garcia2013c2c}. Adopting these metrics, we define a new metric for JPMS based projects, which is called the module-to-module ($M2M$) metric.
Our metric is based on architectural changes at the module ($M$) level, arguably the higher level. Constraints of the M2M metric are defined based on A2A (or C2C) and ID-SD (include and symbol dependency) \cite{lutellier2015archi_recovercompare} metrics suited for JPMS. Deleting, adding and moving modules or their respective classes are considered A2A delta operations. In contrast, ID-SD considers the modification of classes and interfaces importing (for Java and Kotlin) from different modules (representing $M(IDSD)$), not from the same module. Module operations (MO) (described in Section \ref{jpms_op}) update within the \emph{module-info.java} files changes at least the runtime architecture irrespective of changes within the class files. Consequently, we also consider the modification of them for the M2M metric.
Hence, a module-level architectural change metric $M2M$ may contain any of $\Delta a2a$, $M(IDSD)$ and $MO$ changes.
\begin{table}
\footnotesize
	\caption{Observation of directory and naming structures}\label{extraction_info}
		\vspace{-.2cm}
	\begin{center}
		\begin{tabular}{|p{.10in}|p{.37in}|p{2.59in}|} \hline
			 \textbf{SL} &\textbf{Type} & \textbf{Description} \\ \hline 
1&Import & Code location change appears as deletion and addition \\\hline
2&Import & Shrinking and elaborating multiple imports appears as deletion and addition \\\hline
3 &Import & VCS APIs only return the first and last lines as modification of the multiple imports  commented with $\backslash*..$ $*\backslash$ \\\hline
4 &Import & New Java allows importing static method and inner class \\\hline
5 &Import& Kotlin offers static method import without any syntax variation \\\hline
6&Import& JPMS config includes both class name and package name \\\hline
7&Dependent & Methods have subsequent relations and contain relative references\\\hline
8&Directory & Both Kotlin and Java modules have uncommon directory structures \\\hline
9&Directory & Directory name of one module could be similar to sub-directory name of another module\\\hline
10&Directory & Some of the modules have only one root package/directory name\\\hline
11&Directory & Module has submodules \\\hline
12&Naming& Import like directory name such as $application.api$ \\\hline
13&Naming& Similar class name in multiple modules \\\hline
14&Naming& Class and package renaming appears as addition and deletion \\\hline
15&Naming& Class and method name in Kotlin do not have different patterns\\\hline
16&Naming& Module name in module-info.java is different from directory name, and used in ambiguous ways in import\\\hline
\end{tabular}
	\end{center}
	\vspace{-0.4cm}
\end{table}
\par
\subsection{M2M Change Detection Process}
During the manual analysis of the 3647 commits, we notice that DANS properties play a crucial role in determining the M2M instances. We observe various properties and their variations across the commits of the projects. The list of these properties (and types) are presented in Table \ref{extraction_info}. However, the M2M instance can be detected in two ways: (i) comparing ASTs from byte code of two consecutive commits, and (ii) processing the provided information by the APIs and libraries of the VCS \cite{gitpython,pydriller}. In the first technique, a complete code-base for each committed version needs to be compiled into byte code, which mostly requires manual intervention to resolve 3rd party library dependencies. Moreover, intensive computation could be a bottleneck for a normal purpose machine for analyzing changes at the statement levels for large systems. 
Usually, each release of a project may contain hundreds of commits. Thus, AST-based techniques might far exceed the manual analysis's time and efforts for architectural change detection. Therefore, we aim for a lightweight tool based on the second technique.
\par
We have extracted code change information in between two consecutive commits by \textit{git} APIs with the help of GitPython \cite{gitpython}, and PyDriller \cite{pydriller}. It provides string/text of the modified code segments with line numbers, methods and classes. Among the 2,720 manually extracted M2M commits, we have selected 48 commits (5 for each project except mvvmFX) from 10 projects having multiple intentions in the commit description. These 48 samples are used as so called training samples in our experiment for automated tool development with the DANS properties. We have manually analyzed all types of changes of those commits and found that code information returned by the VCS APIs has some non-trivial challenges that might compromise the detection process's accuracy. These challenges are described in Table \ref{extraction_info}. Here, we discuss some of them. For example, for SL8 in the table, we have identified at least three types of directory structures of JPMS modules where the class files might reside in: \emph{"module\_name/\textbf{src}/class\_dir"}, \emph{"module\_name/\textbf{main/java}/class\_dir"}, and \emph{"module\_name/\textbf{src/main/\\java}/class\_dir"}; but, this information is not included within the class imports as can be seen in Figure \ref{example_change}. One instance for SL2 is that \emph{import aa.1, import aa.2, and import aa.3} can be shrink to \emph{import aa.*;} and vice-versa. Another complex challenge is distinguishing between \emph{import renaming} and new \emph{import addition} in SL1. For example, renaming \emph{import aa.\textbf{b}.1} to \emph{aa.\textbf{c}.1} due to directory renaming is appeared as a deletion and an addition operation. One example for SL16 is that the similar directory structure \textit{ch/tutteli/atrium/core/api} of the module name \textit{ch.tutteli.atrium.core.api} does not exist up to that commit but used within other module as import. Comment within the change information also poses challenges, such as commenting as shown in SL3. We point out some other anomalies in the commits and handle all these concerns using regular expressions in string processing. We have developed a tool in the Python platform to handle these observations for the DANS properties.

\section{Semantic Slice Generation}  \label{slicing}
Based on the DANS properties in Table \ref{extraction_info}, we generate semantic summary (architectural) containing relational information (SSC) of the changed code snippets presented in Table \ref{semantic_summmary}. We extract 16 types of change relations (such as cross-module class used in a newly defined method) involved in architectural change instances. One of the slices for Figure \ref{example_change} would be, \emph{ASBC:EBCB$=>$sasToken$<$- ASC:STC}. As discussed in Section \ref{jpms_op}, this slice represents that it is a M2M instance where \emph{EBCB} class of \emph{ASBC} module added \emph{sasToken} method that is dependent on the \emph{STC} class of \emph{ASC} module. A complex change might contain many such slices of all the information in table \ref{semantic_summmary}.
Our extraction process depends on directory processing, \emph{module-info.java} file processing, methods processing, and searching within the programs before and after modification with regular expressions for extracting the DANS properties. Initially, we thought the process would be straight-forward. The challenges described in the previous section significantly influence the performance of the automated technique. However, for including a method within a slice, we handle the SL7 in the table as follows (along with other common concerns such as removing the Java keywords). Let's consider that class A is involved in an M2M instance, then following would be the candidates of the search process for a method:
\begin{itemize}
\small
    \item B $<$T$>$ objectB = new B$<$A$>$(), then objectB is used.
        \item B objectB = C.getObj(A), then objectB is used.
        \item B getObj(A), then getObj is used.
        \item All the cases directly assigned in variables and used in methods.

\end{itemize}
A few of the challenges are compromised due to better performance since resolving those introduces other problems (mostly due to static method import) and worsen the outcome. The extracted information is saved into \emph{yaml} template so that any tool can read the data for further purposes.
\begin{table}
	\caption{Information in the semantic slices}\label{semantic_summmary}
		\vspace{-.2cm}
	\small
	\begin{center}
		\begin{tabular}{|p{.10in}|p{1.10in}|p{1.80in}|} \hline
			 \textbf{\#} &\textbf{Entity} & \textbf{Change Relation} \\ \hline 
            1& JPMS$>>$Direct module/ MO add, delete, modify & connected jpms+API modules \par disconnected jpms+API modules  \\\hline
            2& JPMS$>>$Added class & connected jpms+API modules and their classes, contextual new methods\\\hline
        3& JPMS$>>$Deleted class & disconnected jpms+API modules and their classes, contextual deleted methods \\\hline
        4& JPMS$>>$Modified class & Relation information in both \# 2 and 3\\\hline
        5& JPMS$>>$Modified class & Not involved in M2M \\\hline
		\end{tabular}
	\end{center}
	\vspace{-0.5cm}
\end{table}
	\vspace{-0.1cm}
\section{Performance evaluation}\label{performance_validate}
We compare the outcome of M2M detection and slice generation with the manual collection and measure recall (R) -- quantitative correctness of retrieving the change instances, and precision (P) -- the accuracy rate among the predicted change instances \cite{Goutte2005f1_probabilistic}. First, we run our tool on all 3,647 samples (except 48) (shown in Table \ref{table:dataset}) having structural changes; 2,720 of them contain M2M metric. We measure the precision (P) and recall (R) excluding the 48 training sets. The individual project's performance result is shown in Table \ref{table:dataset} (P and R columns; MvvmFX has no test M2Ms left).
In some cases, performance is compromised for the static import. In the worst case, our tool's precision rate is 97\%, and the recall rate is 96\%. The highest number of incorrect outcomes is for Vooga (12). For many projects, both the P and R are 100\%. Therefore, the DANS property is highly reliable for detecting M2M instances.
\par
For the preliminary investigation of the SSC generation of the M2M instances, we explored the 48 training samples. First, we manually extracted (and saved into YAML files) all the slices of those commits. Then, we evaluate the automated technique's outcome based on the involved entities of a slice with that ground truth. For instance, the discussed slice in Section \ref{slicing} has five entities/instances. If a module itself modifies its dependency with other modules and appears within the other two modules, the instance count would be three; this is true for all other cases. The total number of such instances in each project is shown in Table \ref{review_data}; it also shows the classes that are not involved in M2M (nonM2M). The outcome of the automated tool is also presented in the $P$ and $R$ columns of Table \ref{review_data}. For each project, P is from 93 to 100\%, and R is from 97 to 100\%. Therefore, the DANS properties are also highly reliable for generating the SSCs and can be extracted without compiling each version's code. However, some of the instances are not properly extracted due to SL 3, 4, 5 and 7 in Table \ref{extraction_info}. The lowest precision is for Bach. We have investigated that the module directory structure is unusual for Bach (e.g., the sub-modules are within the \textit{src} folder of the main module), and solving those actually decreases the overall performance significantly. The technique produces the most number of incorrect instances for the AzureSDK (57). The performance is quite general because the investigation is conducted for ten projects with two language frameworks. Our tool cannot process anonymous inner class methods since the \textit{git} API does not provide separate (and structured) information about that.

\begin{table}[h]
	\small
	\caption{Semantic slice data and performance outcome.}\label{review_data}
	\vspace{-.2cm}
	\begin{center}
		\begin{tabular}{|p{.55in}|p{0.38in}|p{.27in} |p{.40in}|p{.20in} |p{.20in} | } \hline
			\textbf{Project} & \textbf{Commit} & \textbf{M2Ms} & \textbf{nonM2M} & \textbf{P} & \textbf{R} \\\hline \hline
			Hibernate  & 5 &220&17& 0.99&1.0\\\hline
			Aion  & 5 & 158&19&1.0&.97\\\hline
		    Webfx  & 5 & 28 &3& 0.94& 1.0\\\hline
			Speedment  & 5 & 278 &22&1.0& 0.98\\\hline
			AzureSDK & 5 & 949& 67& 0.95& 0.99\\\hline 
			Atrium & 5 & 135&36 & 0.99& 0.98\\\hline
			Bach  & 5 &48 &2& 0.93& 1.0\\\hline
		    Vooga  & 5 & 99& 9&1.0&1.0\\\hline
			Imgui  & 5 &52  &10&1.0&0.97\\\hline
			MvvmFX & 3 & 96 &4& 0.98&1.0\\\hline 
		\end{tabular}
	\end{center}
		\vspace{-.4cm}
\end{table}

\textbf{Bias Testing:}
To reduce bias in performance testing for the automated SSC generation, we measure the outcome of our proposed tool with 16 unseen commits (two samples from each of eight projects). Those samples are randomly selected, excluding the experimental set (48 commits), and the SSCs are first manually extracted. Then the detected slices are compared against the manual extracted set. The performance outcome of our tool is shown in Table \ref{bias_test}. The lowest precision rate with this dataset is 91\%, and the lowest recall rate is 96\%. Therefore, the bias testing also confirms the performance of our tool for semantic change slice generation.

\begin{table}[h]
	\small
	\caption{Bias testing outcome.}\label{bias_test}
	\vspace{-.2cm}
	\begin{center}
		\begin{tabular}{|p{.55in}|p{0.38in}|p{.27in} |p{.40in}|p{.20in} |p{.20in} | } \hline
			\textbf{Project} & \textbf{Commit} & \textbf{M2Ms} & \textbf{nonM2M} & \textbf{P} & \textbf{R} \\\hline \hline
			Aion  & 2 & 114&9&0.91&1.0\\\hline
		    Webfx  & 2 & 651 &39& 0.96& 0.98\\\hline
			Speedment  & 2 & 124 &19&1.0& 0.98\\\hline
			AzureSDK & 2 & 79& 9& 1.0& 1.0\\\hline 
			Atrium & 2 & 106&6 & 1.0& 1.0\\\hline
			Bach  & 2 &21 &4& 0.98& 0.99\\\hline
		    Vooga  & 2 & 50& 6&1.0&0.96\\\hline
			Imgui  & 2 &418  &85&0.98&0.99\\\hline
		\end{tabular}
	\end{center}
		\vspace{-.4cm}
\end{table}
\section{Conclusion and Future Work}
In this paper, we present our initial observation on the impact of DANS properties to develop a design review tool that detects and semantically slices the architectural change instances of a commit. Performance evaluation with ten open-source projects proves that this process produces reliable outcomes while the technique is lightweight. We will cover more concerns in the tool, such as extracting indirectly impacted methods that invoke methods in Table \ref{semantic_summmary}. We believe that the directory-based challenges that we have discussed in this paper will persist in the AST-based approaches (since AST nodes are generated from the directory structure information). Furthermore, we will evaluate the performance of various types of slices presented in this table. Semantic change information presented in Table \ref{semantic_summmary} can be utilized to generate more understandable code descriptions \cite{jiang2017change_summary} and can be mapped with multiple intentions if they exist within a commit. Our tool would be useful for a number of empirical studies besides assisting design review, such as the effectiveness measures of the existing design decision recovery approaches, determining architectural change types, developers profile buildup based on design changes, design debt and change impact analysis, release note generation, design change versioning scheme, etc. Dataset and the script of the tool are available \cite{archslice} for further advancement.
\par
\textbf{Future Work:}
Our main objective is to assist in semantic design review. Semantic design recovery and semantic design summary (for each release or milestone) generations are the essential steps for that. For that purpose, we plan to investigate concept generation by mapping with the commit description and code identifiers associated with each of the DANS properties within the change instances with our proposed tool. Semantic software design is involved in the concept/meaning of software features/requirements associating design logics (including architecture) and implementation in the programming languages \cite{hewitt2019semantic}. That is why semantic slicing is essential for semantic design recovery.
The architectural change relations and concept generation would facilitate to advance of our planned empirical study on semantic design recovery and summary generation. We also explore separating tangled commits (having M2M) with DANS properties, which is also required for the efficient design review tool. Moreover, our proposed tool needs to be enhanced in string pattern matching as we have observed that in some cases, almost identical directory structures are falsely identified (completely ignoring them reduces the tool performance significantly). To handle this situation, we will experiment with the \emph{Context Triggered Piecewise Hash} \cite{kornblum2006identifying} mechanism.

\section*{Acknowledgment}
\footnotesize
This research is supported by the Natural Sciences and Engineering Research Council of Canada (NSERC), and by two Canada First Research Excellence Fund (CFREF) grants coordinated by the Global Institute for Food Security (GIFS) and the Global Institute for WaterSecurity (GIWS).
\bibliographystyle{ACM-Reference-Format}
\bibliography{esem21}

\end{document}